\documentclass[fleqn,11pt]{article}

\usepackage{latexsym,ifthen,epsfig,color}
\usepackage{amsmath,amssymb,amsthm}
\usepackage{float}

\newcommand{\join}{\text{\textcircled{{\footnotesize 1}}}}
\newcommand{\cojoin}{\text{\textcircled{{\footnotesize 0}}}}

\newcommand{\NP}{\ensuremath{\mathbb{NP}}}

\newtheorem{theorem}{Theorem}
\newtheorem{lemma}{Lemma}

\newtheorem{corollary}{Corollary}
\newtheorem{proposition}{Proposition}

\newtheorem{clai}{Claim}
\newtheorem{observation}{Observation}

\begin{document}

\title{Finding Dominating Induced Matchings in $P_9$-Free Graphs in Polynomial Time}

\author{
Andreas Brandst\"adt\footnote{Institut f\"ur Informatik,
Universit\"at Rostock, A.-Einstein-Str.\ 22, D-18051 Rostock, Germany,
{\texttt andreas.brandstaedt@uni-rostock.de}}
\and
Raffaele Mosca\footnote{Dipartimento di Economia, Universit\'a degli Studi ``G.\ D'Annunzio''
Pescara 65121, Italy.
{\texttt r.mosca@unich.it}}
}

\maketitle


\begin{abstract}
Let $G=(V,E)$ be a finite undirected graph. An edge subset $E' \subseteq E$ is a {\em dominating induced matching} ({\em d.i.m.}) in $G$ if every edge in $E$ is intersected by exactly one edge of $E'$.
The \emph{Dominating Induced Matching} (\emph{DIM}) problem asks for the existence of a d.i.m.\ in $G$.
The DIM problem is \NP-complete even for very restricted graph classes such as planar bipartite graphs with maximum degree 3 but was solved in linear time for $P_7$-free graphs and in polynomial time for $P_8$-free graphs.
In this paper, we solve it in polynomial time for $P_9$-free graphs.
\end{abstract}


\section{Introduction}\label{sec:intro}

Let $G=(V,E)$ be a finite simple undirected graph, i.e., an undirected graph without loops and multiple edges. Given an edge $e \in E$, we say that $e$ {\em dominates} itself and every edge sharing a vertex with $e$.
An edge subset $M \subseteq E$ is an {\em induced matching} if the pairwise distance between its members is at least 2 (i.e., the {\em distance property}), that is, $M$ is isomorphic to $kP_2$ for $k=|M|$. A subset $M \subseteq E$ is a {\em dominating induced matching} ({\em d.i.m.} for short) of $G$ if $M$ is an induced matching in $G$ such that every edge in $E$ is dominated by exactly one edge in $M$. Clearly, not every graph $G$ has a d.i.m.; the {\sc Dominating Induced Matching} (DIM) problem asks for the existence of a d.i.m.\ in $G$.

The DIM problem is also called {\sc Efficient Edge Domination} (EED) in various papers:
Recall that a vertex $v \in V$ {\em dominates} itself and its neighbors. A vertex subset $D \subseteq V$ is an {\em efficient dominating set} ({\em e.d.s.} for short) of $G$ if every vertex of $G$ is dominated by exactly one vertex in $D$.
The notion of efficient domination was introduced by Biggs \cite{Biggs1973} under the name {\em perfect code}.
The {\sc Efficient Domination} (ED) problem asks for the existence of an e.d.s.\ in a given graph $G$ (note that not every graph has an e.d.s.)
A set $M$ of edges in a graph $G$ is an \emph{efficient edge dominating set} (\emph{e.e.d.s.} for short) of $G$ if and only if it is an e.d.s.\ in its line graph $L(G)$. The {\sc Efficient Edge Domination} (EED) problem asks for the existence of an e.e.d.s.\ in a given graph $G$. Thus, the EED problem for a graph $G$ corresponds to the ED problem for its line graph $L(G)$. Note that not every graph has an e.e.d.s.

In \cite{GriSlaSheHol1993}, it was shown that the DIM problem is \NP-complete; see also~\cite{BraHunNev2010,CarKorLoz2011,LuKoTan2002,LuTan1998}.
However, for various graph classes, DIM is solvable in polynomial time. For mentioning some examples, we need the following notions:

Let $P_k$ denote the chordless path $P$ with $k$ vertices, say $a_1,\ldots,a_k$, and $k-1$ edges $a_ia_{i+1}$, $1 \le i \le k-1$; we also denote it as $P=(a_1,\ldots,a_k)$.

For indices $i,j,k \ge 0$, let $S_{i,j,k}$ denote the graph $H$ with vertices $u,x_1,\ldots,x_i$, $y_1,\ldots,y_j$, $z_1,\ldots,z_k$ such that the subgraph induced by $u,x_1,\ldots,x_i$ forms a $P_{i+1}$ $(u,x_1,\ldots,x_i)$, the subgraph induced by $u,y_1,\ldots,y_j$ forms a $P_{j+1}$ $(u,y_1,\ldots,y_j)$, and the subgraph induced by $u,z_1,\ldots,z_k$ forms a $P_{k+1}$ $(u,z_1,\ldots,z_k)$, and there are no other edges in $S_{i,j,k}$; $u$ is called the {\em center} of $H$.
Thus, {\em claw} is $S_{1,1,1}$, and $P_k$ is isomorphic to $S_{k-1,0,0}$.

For a set ${\cal F}$ of graphs, a graph $G$ is called {\em ${\cal F}$-free} if no induced subgraph of $G$ is contained in ${\cal F}$.
If $|{\cal F}|=1$, say ${\cal F}=\{H\}$, then instead of $\{H\}$-free, $G$ is called $H$-free.

\medskip

The following results are known:

\begin{theorem}\label{DIMpolresults}
DIM is solvable in polynomial time for
\begin{itemize}
\item[$(i)$]  $S_{1,1,1}$-free graphs $\cite{CarKorLoz2011}$,
\item[$(ii)$] $S_{1,2,3}$-free graphs $\cite{KorLozPur2014}$,
\item[$(iii)$] $S_{2,2,2}$-free graphs $\cite{HerLozRieZamdeW2015}$,
\item[$(iv)$] $S_{1,2,4}$-free graphs $\cite{BraMos2017/2}$,
\item[$(v)$] $S_{2,2,3}$-free graphs $\cite{BraMos2017/3}$,
\item[$(vi)$] $S_{1,1,5}$-free graphs $\cite{BraMos2019}$,
\item[$(vii)$]  $P_7$-free graphs $\cite{BraMos2014}$ (in this case even in linear time),
\item[$(viii)$] $P_8$-free graphs $\cite{BraMos2017}$.
\end{itemize}
\end{theorem}

In \cite{HerLozRieZamdeW2015}, it is conjectured that for every fixed $i,j,k$, DIM is solvable in polynomial time for $S_{i,j,k}$-free graphs (actually, an even stronger conjecture is mentioned in \cite{HerLozRieZamdeW2015}); this includes $P_k$-free graphs for $k \ge 9$.
In this paper we show that DIM can be solved in polynomial time for $P_9$-free graphs (generalizing the corresponding results for $P_7$-free and for $P_8$-free graphs).

\section{Definitions and Basic Properties}\label{sec:basicnotionsresults}

\subsection{Basic notions}\label{subsec:basicnotions}

Let $G$ be a finite undirected graph without loops and multiple edges. Let $V(G)$ or $V$ denote its vertex set and $E(G)$ or $E$ its edge set (say $G=(V,E)$);
let $n=|V|$ and $m=|E|$.
For $v \in V$, let $N(v):=\{u \in V: uv \in E\}$ denote the {\em open neighborhood of $v$}, and let $N[v]:=N(v) \cup \{v\}$ denote the {\em closed neighborhood of $v$}. If $xy \in E$, we also say that $x$ and $y$ {\em see each other}, and if $xy \not\in E$, we say that $x$ and $y$ {\em miss each other}. A vertex set $S$ is {\em independent} in $G$ if for every pair of vertices $x,y \in S$, $xy \not\in E$. A vertex set $Q$ is a {\em clique} in $G$ if for every pair of vertices $x,y \in Q$, $x \neq y$, $xy \in E$. For $uv \in E$ let $N(uv):= N(u) \cup N(v) \setminus \{u,v\}$ and $N[uv]:= N[u] \cup N[v]$.

For $U \subseteq V$, let $G[U]$ denote the subgraph of $G$ induced by vertex set $U$. Clearly $xy \in E$ is an edge in $G[U]$ exactly when $x \in U$ and $y \in U$; thus, $G[U]$ can simply be denoted by $U$ (if understandable).

For $A \subseteq V$ and $B \subseteq V$, $A \cap B = \emptyset$, we say that $A \cojoin B$ ($A$ and $B$ {\em miss each other}) if there is no edge between $A$ and $B$, and $A$ and $B$ {\em see each other} if there is at least one edge between $A$ and $B$. If a vertex $u \notin B$ has a neighbor $v \in B$ then {\em $u$ contacts $B$}. If every vertex in $A$ sees every vertex in $B$, we denote it by $A \join B$. For $A=\{a\}$, we simply denote  $A \join B$ by $a \join B$, and correspondingly $A \cojoin B$ by $a \cojoin B$.
If for $A' \subseteq A$, $A' \cojoin (A \setminus A')$, we say that $A'$ is {\em isolated} in $G[A]$.
For graphs $H_1$, $H_2$ with disjoint vertex sets, $H_1+H_2$ denotes the disjoint union of $H_1$, $H_2$, and for $k \ge 2$, $kH$ denotes the disjoint union of $k$ copies of $H$. For example, $2P_2$ is the disjoint union of two edges.

As already mentioned, a {\em chordless path} $P_k$, $k \ge 2$, has $k$ vertices, say $v_1,\ldots,v_k$, and $k-1$ edges $v_iv_{i+1}$, $1 \le i \le k-1$;
the {\em length of $P_k$} is $k-1$. We also denote it as $P=(v_1,\ldots,v_k)$.

A {\em chordless cycle} $C_k$, $k \ge 3$, has $k$ vertices, say $v_1,\ldots,v_k$, and $k$ edges $v_iv_{i+1}$, $1 \le i \le k-1$, and $v_kv_1$; the {\em length of $C_k$} is $k$.

Let $K_i$, $i \ge 1$, denote the clique with $i$ vertices. Let $K_4-e$ or {\em diamond} be the graph with four vertices, say $v_1,v_2,v_3,u$, such that $(v_1,v_2,v_3)$ forms a $P_3$ and $u \join \{v_1,v_2,v_3\}$; its {\em mid-edge} is the edge $uv_2$.

A {\em butterfly} has five vertices, say, $v_1,v_2,v_3,v_4,u$, such that $v_1,v_2,v_3,v_4$ induce a $2P_2$ with edges $v_1v_2$ and $v_3v_4$ (the {\em peripheral edges} of the butterfly), and $u \join \{v_1,v_2,v_3,v_4\}$.

We often consider an edge $e = uv$ to be a set of two vertices; then it makes sense to say, for example, $u \in e$ and $e \cap e' \neq \emptyset$, for an edge $e'$. For two vertices $x,y \in V$, let $dist_G(x,y)$ denote the {\em distance between $x$ and $y$ in $G$}, i.e., the length of a shortest path between $x$ and $y$ in $G$.
The {\em distance between a vertex $z$ and an edge $xy$} is the length of a shortest path between $z$ and $x,y$, i.e., $dist_G(z,xy)= \min\{dist_G(z,v): v \in \{x,y\}\}$.
The {\em distance between two edges} $e,e' \in E$ is the length of a shortest path between $e$ and $e'$, i.e., $dist_G(e,e')= \min\{dist_G(u,v): u \in e, v \in e'\}$.
In particular, this means that $dist_G(e,e')=0$ if and only if $e \cap e' \neq \emptyset$.

Clearly, $G$ has a d.i.m.\ if and only if every connected component of $G$ has a d.i.m.; from now on, we consider
that $G$ is connected, and connected components of induced subgraphs of $G$ are mentioned as {\em components}.

Note that if $G=(V,E)$ has a d.i.m.\ $M$, and $V(M)$ denotes the vertex set of $M$ then $V \setminus V(M)$ is an independent set, say $I$, i.e.,
\begin{equation}\label{IV(M)partition}
V \mbox{ has the partition } V = V(M) \cup I .
\end{equation}

From now on, all vertices in $I$ are colored white and all vertices in $V(M)$ are colored black. According to \cite{HerLozRieZamdeW2015}, we also use the following notions: A partial black-white coloring of $V$ is {\em feasible} if the set of white vertices is an independent set in $G$ and every black vertex has at most one black neighbor. A complete black-white coloring of $V$ is {\em feasible} if the set of white vertices is an independent set in $G$ and every black vertex has exactly one black neighbor. Clearly, $M$ is a d.i.m.\ of $G$ if and only if the black vertices $V(M)$ and the white vertices $V \setminus V(M)$ form a complete feasible coloring of $V$.

\subsection{Reduction steps, forbidden subgraphs, forced edges, and excluded edges}\label{forbidsubgrforcededges}

Various papers on this topic introduced and applied some {\em forcing rules} for reducing the graph $G$ to a subgraph $G'$ such that $G$ has a d.i.m.\ if and only if $G'$ has a d.i.m., based on the condition that for a d.i.m.\ $M$, $V$ has the partition $V= V(M) \cup I$ such that all vertices in $V(M)$ are black and all vertices in $I$ are white (recall (\ref{IV(M)partition})).

\medskip

A vertex $v \in V$ is {\em forced to be black} if for every d.i.m.\ $M$ of $G$, $v \in V(M)$.
Analogously, a vertex $v \in V$ is {\em forced to be white} if for every d.i.m.\ $M$ of $G$, $v \notin V(M)$.

Clearly, if $uv \in E$ and if $u,v$ are forced to be black, then $uv$ is contained in every (possible) d.i.m. of $G$.

\medskip

An edge $e \in E$ is a {\em forced} edge of $G$ if for every d.i.m.\ $M$ of $G$,
$e \in M$. Analogously, an edge $e \in E$ is an {\em excluded} edge of $G$ if for every d.i.m.\ $M$ of $G$,
$e \not \in M$.

\medskip

For the correctness of the reduction steps, we have to argue that $G$ has a d.i.m.\ if and only if the reduced graph $G'$ has one (provided that no contradiction arises in the vertex coloring, i.e., it is feasible).

Then let us introduce two reduction steps which will be applied later.

\medskip

\noindent
{\bf Vertex Reduction.} Let $u \in V(G)$. If $u$ is forced to be white, then
\begin{itemize}
\item[$(i)$] color black all neighbors of $u$, and
\item[$(ii)$] remove $u$ from $G$.
\end{itemize}

Let $G'$ be the reduced subgraph. Clearly, Vertex Reduction is correct, i.e., $G$ has a d.i.m.\ if and only if $G'$ has a d.i.m. \\

\noindent
{\bf Edge Reduction.} Let $uv \in E(G)$. If $u$ and $v$ are forced to be black, then
\begin{itemize}
\item[$(i)$] color white all neighbors of $u$ and of $v$ (other than $u$ and $v$), and
\item[$(ii)$] remove $u$ and $v$ (and the edges containing $u$ or $v$) from $G$.
\end{itemize}

Again, clearly, Edge Reduction is correct, i.e., $G$ has a d.i.m.\ if and only if the reduced subgraph $G'$ has a d.i.m. \\

The subsequent notions and observations lead to some possible reductions (some of them are mentioned e.g.\ in \cite{BraHunNev2010,BraMos2014,BraMos2017}).

\begin{observation}[\cite{BraHunNev2010,BraMos2014,BraMos2017}]\label{dimC3C5C7C4}
Let $M$ be a d.i.m.\ of $G$.
\begin{itemize}
\item[$(i)$] $M$ contains at least one edge of every odd cycle $C_{2k+1}$ in $G$, $k \ge 1$,
and exactly one edge of every odd cycle $C_3$, $C_5$, $C_7$ in $G$.
\item[$(ii)$] No edge of any $C_4$ can be in $M$.
\item[$(iii)$] For each $C_6$ either exactly two or none of its edges are in $M$.
\end{itemize}
\end{observation}

\noindent
{\bf Proof.} See e.g.\ Observation 2 in \cite{BraMos2014}.

\medskip

In what follows, we will also refer to Observation \ref{dimC3C5C7C4}~$(i)$ (with respect to $C_3$) as to the {\em triangle-property}, and to Observation \ref{dimC3C5C7C4}~$(ii)$ as to the {\em $C_4$-property}.

\medskip

Since by Observation \ref{dimC3C5C7C4} $(i)$, every triangle contains exactly one $M$-edge, and the pairwise distance of $M$-edges is at least 2, we have:

\begin{corollary}\label{cly:K4free}
If $G$ has a d.i.m.\ then $G$ is $K_4$-free.
\end{corollary}

\noindent
{\sc Assumption 1.} From now on, by Corollary \ref{cly:K4free}, we assume that the input graph is $K_4$-free (else it has no d.i.m.).

\medskip

Clearly, it can be checked (directly) in polynomial time whether the input graph is $K_4$-free.

\medskip

By Observation \ref{dimC3C5C7C4} $(i)$ with respect to $C_3$ and the distance property, we have the following:

\begin{observation}\label{obs:diamondbutterfly}
The mid-edge of any diamond in $G$ and the two peripheral edges of any induced butterfly are forced edges of $G$.
\end{observation}

\noindent
{\sc Assumption 2.} From now on, by Observation \ref{obs:diamondbutterfly}, we assume that the input graph is (diamond,butterfly)-free.

\medskip

In particular, we can apply the Edge Reduction to each mid-edge of any induced diamond and to each peripheral edge of any induced butterfly; that can be done in polynomial time.

\medskip

Here is an example for excluded edges:
By Observation \ref{dimC3C5C7C4} $(i)$, there is exactly one $M$-edge in the $C_3$ $(v_1,v_2,v_3)$. Since $G$ is $K_4$- and diamond-free, every vertex $v \notin \{v_1,v_2,v_3\}$ which contacts the $C_3$ $(v_1,v_2,v_3)$ has exactly one neighbor in $(v_1,v_2,v_3)$.

A {\em paw} has four vertices, say $v_1,v_2,v_3,v_4$ such that $v_1,v_2,v_3$ induce a $C_3$ and $v_4$ contacts exactly one vertex in $v_1,v_2,v_3$, say $v_3v_4 \in E$. Thus, the edge $v_3v_4 \in E$ is excluded.

\subsection{The distance levels of an $M$-edge $xy$ in a $P_3$}\label{subsec:distlevels}

Based on \cite{BraMos2017}, we first describe some general structure properties for the distance levels of an edge in a d.i.m.\ $M$ of $G$.
Since $G$ is $(K_4$,diamond,butterfly)-free, we have:

\begin{observation}\label{obse:neighborhood}
For every vertex $v$ of $G$, $N(v)$ is the disjoint union of isolated vertices and at most one edge. Moreover, for every edge $uv \in E$, there is at most one common neighbor of $u$ and $v$.
\end{observation}

Since it is trivial to check whether $G$ has a d.i.m.\ $M$ with exactly one edge, from now on we can assume that $|M| \geq 2$.

\medskip

Recall that the distance $dist_G(a,b)$ between two vertices $a,b$ in graph $G$ is the number of edges in a shortest path in $G$ between $a$ and $b$.

\begin{theorem}[\cite{BacTuz1990}]\label{theoB-T}
Every connected $P_t$-free graph $G=(V,E)$ admits a vertex $v \in V$ such that $dist_G(v,w) \leq \bigl\lfloor t/2 \bigr\rfloor$ for every $w \in V$.
\end{theorem}

We call such a vertex $v$ a {\em central} vertex; more exactly, a central vertex in $G$ has shortest distance to every other vertex in $G$.
Theorem \ref{theoB-T} implies that every connected $P_9$-free graph $G$ admits a central vertex $v \in V$ such that $dist_G(v,w) \leq 4$ for every $w \in V$. For a central vertex $v$ and a neighbor $u$ of $v$, i.e., $uv \in E$, let
$$N_i(uv):=\{z \in V: dist_G(z,uv) = i\}$$
denote the {\em distance levels of $uv$}, $i \ge 1$. Then by Theorem \ref{theoB-T}, for every edge $uv \in E$, we have
\begin{equation}\label{N5empty}
N_k(uv)=\emptyset \mbox{ for every } k \ge 5.
\end{equation}

\begin{observation}\label{obse:centralvertexinP3}
For every central vertex $v$ in $G$, every edge $uv \in E$ is part of a $P_3$ of $G$.
\end{observation}

\noindent
{\bf Proof.}
Let $v$ be a central vertex in $G$, and suppose to the contrary that not every edge $uv \in E$ is part of a $P_3$ of $G$, say $(u,v,w)$ induce a $C_3$ in $G$ such that $uv$ is not part of a $P_3$, i.e., $N[u]=N[v]=\{u,v,w\}$. Clearly, in this case, $w$ has more neighbors than $v$ in $G$ since $G$ itself is no $C_3$, i.e.,
$w$ has a neighbor $x$ with $xu \notin E$ and $xv \notin E$ (else there is a diamond or $K_4$). Moreover, $w$ is not part of a triangle with $x$ and $y$ (else there is a butterfly in $G$).
Then for every vertex $y \notin \{u,v,w\}$, $dist_G(w,y) < dist_G(v,y)$, which is a contradiction.

\medskip

Thus, Observation \ref{obse:centralvertexinP3} is shown.
\qed

\medskip

Now assume that $v$ is a central vertex in $G$ such that every edge $uv \in E$ is part of a $P_3$ of $G$.
Then one could check for any edge $uv \in E$ (with central vertex $v$), whether there is a d.i.m.\ $M$ of $G$ with $uv \in M$, and
one could conclude: Either $G$ has a d.i.m.\ $M$ with $v \in V(M)$, or $G$ has no d.i.m.\ $M$ with $v \in V(M)$; in particular, in the latter case, if none of the edges $uv$ is in a d.i.m.\ then $v$ is white and one can apply the Vertex Reduction to $v$ and in particular remove $v$.

\medskip

Now assume that $x$ is a central vertex (as in Observation \ref{obse:centralvertexinP3}), and let $xy \in M$ be an $M$-edge for which there is a vertex $r$ such that $\{r,x,y\}$ induce a $P_3$ with edge $rx \in E$. By the assumption that $xy \in M$, we have that $x$ and $y$ are black, and it could lead to a feasible $xy$-coloring (if no contradiction arises).

\medskip

Let $N_0(xy):=\{x,y\}$ and for $i \ge 1$, let $$N_i(xy):=\{z \in V: dist_G(z,xy) = i\}$$ denote the {\em distance levels of $xy$}.
Recall (\ref{N5empty}) which also shows that $N_5(xy)=\emptyset$.
We consider a partition of $V$ into $N_i=N_i(xy)$, $0 \le i \le 4$, with respect to the edge $xy$ (under the assumption that $xy \in M$).

\begin{observation}\label{N4P6}
If $v \in N_i$ for $i \ge 4$ then $v$ is an endpoint of an induced $P_6$, say with vertices $v,v_1,v_2,v_3,v_4,v_5$ such that $v_1,v_2,v_3,v_4,v_5 \in \{x,y\} \cup N_1 \cup \ldots \cup N_{i-1}$ and with edges $vv_1 \in E$, $v_1v_2 \in E$, $v_2v_3 \in E$, $v_3v_4 \in E$, $v_4v_5 \in E$. Analogously, if $v \in N_3$ then $v$ is an endpoint of a corresponding induced $P_5$.
\end{observation}

\noindent
{\bf Proof.} If $i \ge 5$ then clearly there is such a $P_6$. Thus, assume that $v \in N_4$. Then $v_1 \in N_3$ and $v_2 \in N_2$.
Recall that $y,x,r$ induce a $P_3$. If $v_2r \in E$ then $v,v_1,v_2,r,x,y$ induce a $P_6$. Thus assume that $v_2r \notin E$. Let $v_3 \in N_1$ be a
neighbor of $v_2$. Now, if $v_3x \in E$ then $v,v_1,v_2,v_3,x,r$ induce a $P_6$, and if  $v_3x \notin E$ but $v_3y \in E$ then $v,v_1,v_2,v_3,y,x$ induce a $P_6$.
Analogously, if $v \in N_3$ then $v$ is an endpoint of an induced $P_5$ (which could be part of the $P_6$ above).
Thus, Observation \ref{N4P6} is shown.
\qed

\medskip

Recall that by (\ref{IV(M)partition}), $V=V(M) \cup I $ is a partition of $V$ where $V(M)$ is the set of black vertices and
$I$ is the set of white vertices which is independent.

Since we assume that $xy \in M$ (and is an edge in a $P_3$), clearly, $N_1 \subseteq I$ and thus:
\begin{equation}\label{N1subI}
N_1 \mbox{ is an independent set of white vertices.}
\end{equation}

Moreover, no edge between $N_1$ and $N_2$ is in $M$. Since $N_1 \subseteq I$ and all neighbors of vertices in $I$ are in $V(M)$, we have:
\begin{equation}\label{N2M2S2}
G[N_2] \mbox{ is the disjoint union of edges and isolated vertices. }
\end{equation}

Let $M_2$ denote the set of edges $uv \in E$ with $u,v \in N_2$ and let $S_2 = \{u_1,\ldots,u_k\}$ denote the set of isolated vertices in $N_2$; $N_2=V(M_2) \cup S_2$ is a partition of $N_2$. Obviously:
\begin{equation}\label{M2subM}
M_2 \subseteq M \mbox{ and } S_2 \subseteq V(M).
\end{equation}

If for $xy \in E$, an edge $e \in E$ is contained in {\bf every} d.i.m.\ $M$ of $G$ with $xy \in M$, we say that $e$ is an {\em $xy$-forced} $M$-edge, and analogously, if an edge $e \in E$ is contained in {\bf no} d.i.m.\ $M$ of $G$ with $xy \in M$, we say that $e$ is {\em $xy$-excluded}. The Edge Reduction for forced edges can also be applied for $xy$-forced edges (then, in the unsuccessful case, $G$ has no d.i.m.\ containing $xy$), and correspondingly for $xy$-forced white vertices (resulting from the black color of $x$ and $y$), the Vertex Reduction can be applied.

\medskip

Obviously, by (\ref{M2subM}), we have:
\begin{equation}\label{M2xymandatory}
\mbox{Every edge in } M_2 \mbox{ is an $xy$-forced $M$-edge}.
\end{equation}

Thus, from now on, after applying the Edge Reduction for $M_2$-edges, we can assume that $V(M_2)=\emptyset$, i.e., $N_2=S_2 = \{u_1,\ldots,u_k\}$. For every $i \in \{1,\ldots,k\}$, let $u'_i \in N_3$ denote the {\em $M$-mate} of $u_i$ (i.e., $u_iu'_i \in M$). Let $M_3=\{u_iu'_i: 1 \le i \le k\}$ denote the set of $M$-edges with one endpoint in $S_2$ (and the other endpoint in $N_3$). Obviously, by (\ref{M2subM}) and the distance condition for a d.i.m.\ $M$, the following holds:
\begin{equation}\label{noMedgesN3N4}
\mbox{ No edge with both ends in } N_3 \mbox{ and no edge between } N_3 \mbox{ and } N_4 \mbox{ is in } M.
\end{equation}

As a consequence of (\ref{noMedgesN3N4}) and the fact that every triangle contains exactly one $M$-edge (recall Observation~\ref{dimC3C5C7C4} $(i)$), we have:
\begin{equation}\label{triangleaN3bcN4}
\mbox{For every $C_3$ $abc$} \mbox{ with } a \in N_3, \mbox{ and } b,c \in N_4, \mbox{ $bc \in M$ is an $xy$-forced $M$-edge}.
\end{equation}

This means that for the edge $bc$, the Edge Reduction can be applied, and from now on, we can assume that there is no such triangle $abc$ with $a \in N_3$ and $b,c \in N_4$, i.e., for every edge $uv \in E$ in $N_4$:
\begin{equation}\label{edgeN4N3neighb}
N(u) \cap N(v) \cap N_3 = \emptyset.
\end{equation}

\medskip

According to $(\ref{M2subM})$ and the assumption that $V(M_2)=\emptyset$ (recall $N_2 = \{u_1,\ldots,u_k\}$), let:
\begin{enumerate}
\item[ ] $T_{one} := \{t \in N_3: |N(t) \cap N_2| = 1\}$,

\item[ ] $T_i := T_{one} \cap N(u_i)$, $1 \le i \le k$, and

\item[ ] $S_3 := N_3 \setminus T_{one}$.
\end{enumerate}

By definition, $T_i$ is the set of {\em private} neighbors of $u_i \in N_2$ in $N_3$ (note that $u'_i \in T_i$),
$T_1 \cup \ldots \cup T_k$ is a partition of $T_{one}$, and $T_{one} \cup S_3$ is a partition of~$N_3$.

\begin{lemma}[\cite{BraMos2017}]\label{lemm:structure2}
The following statements hold:
\begin{enumerate}
\item[$(i)$] For all $i \in \{1,\ldots,k\}$, $T_i \cap V(M)=\{u_i'\}$.
\item[$(ii)$] For all $i \in \{1,\ldots,k\}$, $T_i$ is the disjoint union of vertices and at most one edge.
\item[$(iii)$] $G[N_3]$ is bipartite.
\item[$(iv)$] $S_3 \subseteq I$, i.e., $S_3$ is an independent subset of white vertices.
\item[$(v)$] If a vertex $t_i \in T_i$ sees two vertices in $T_j$, $i \neq j$, $i,j \in \{1,\ldots,k\}$, then $u_it_i \in M$ is an $xy$-forced $M$-edge.
\end{enumerate}
\end{lemma}

\noindent
{\bf Proof.} $(i)$: Holds by definition of $T_i$ and by the distance condition of a d.i.m.\ $M$.

\noindent
$(ii)$: Holds by Observation \ref{obse:neighborhood}.

\noindent
$(iii)$: Follows by Observation \ref{dimC3C5C7C4} $(i)$ since every odd cycle in $G$ must contain at least one $M$-edge, and by (\ref{noMedgesN3N4}).

\noindent
$(iv)$: If $v \in S_3:= N_3 \setminus T_{one}$, i.e., $v$ sees at least two $M$-vertices then clearly, $v \in I$, and thus, $S_3 \subseteq I$ is an independent subset (recall that $I$ is an independent set).

\noindent
$(v)$: Suppose that $t_1 \in T_1$ sees $a$ and $b$ in $T_2$. If $ab \in E$ then $u_2,a,b,t_1$ would induce a diamond in $G$. Thus, $ab \notin E$ and now,
$u_2,a,b,t_1$ induce a $C_4$ in $G$; by Observation \ref{dimC3C5C7C4} $(ii)$, no edge in the $C_4$ is in $M$, and by (\ref{noMedgesN3N4}), the only possible $M$-edge for dominating $t_1a,t_1b$ is $u_1t_1$, i.e., $t_1=u'_1$.
\qed

\medskip

By Lemma \ref{lemm:structure2} $(iv)$ and the Vertex Reduction for the white vertices of $S_3$, we can assume:

\begin{itemize}
\item[(A1)] $S_3=\emptyset$, i.e., $N_3=T_1 \cup \ldots \cup T_k$.
\end{itemize}

By Lemma \ref{lemm:structure2} $(v)$, we can assume:

\begin{itemize}
\item[(A2)] For $i,j \in \{1,\ldots,k\}$, $i \neq j$, every vertex $t_i \in T_i$ has at most one neighbor in $T_j$.
\end{itemize}

In particular, if for some $i \in \{1,\ldots,k\}$, $T_i=\emptyset$, then there is no d.i.m.\ $M$ of $G$ with $xy \in M$, and if $|T_i|=1$, say $T_i=\{t_i\}$, then $u_it_i$ is an $xy$-forced $M$-edge.
Thus, we can assume:

\begin{itemize}
\item[(A3)] For every $i \in \{1,\ldots,k\}$, $|T_i| \ge 2$.
\end{itemize}

Let us say that a vertex $t \in T_i$, $1 \le i \le k$, is an {\em out-vertex} of $T_i$ if it is adjacent to some vertex of $T_j$ with $j \neq i$,
or it is adjacent to some vertex of $N_4$, and $t$ is an {\em in-vertex} of $T_i$ otherwise.

For finding a d.i.m.\ $M$ with $xy \in M$, one can remove all but one in-vertices; that can be done in polynomial time.
In particular, if there is an edge between two in-vertices $t_1t_2 \in E$,  $t_1,t_2 \in T_i$, then either $t_1$ or $t_2$ is black, and thus, $T_i$ is completely colored. Thus, let us assume:

\begin{itemize}
\item[(A4)] For every $i \in \{1,\ldots,k\}$, $T_i$ has at most one in-vertex.
\end{itemize}

\begin{lemma}\label{S115fr3edgesbetweenTiTj}
Assume that $G$ has a d.i.m.\ $M$ with $xy \in M$. Then:
\begin{itemize}
\item[$(i)$] For every $i \neq j$, there are at most two edges between $T_i$ and $T_j$.
\item[$(ii)$] If there are two edges between $T_i$ and $T_j$, say $t_it_j \in E$ and $t'_it'_j \in E$ for $t_i,t'_i \in T_i$ and $t_j,t'_j \in T_j$,
$t_i \neq t'_i$, $t_j \neq t'_j$, then every vertex in $(T_i \cup T_j) \setminus \{t_i,t_j,t'_i,t'_j\}$ is white.
\end{itemize}
\end{lemma}

\noindent
{\bf Proof.}
$(i)$: Suppose to the contrary that there are three edges between $T_1$ and $T_2$, say $t_1t_2 \in E$, $t'_1t'_2 \in E$, and $t''_1t''_2 \in E$ for
$t_i,t'_i,t''_i \in T_i$, $i=1,2$. By (A2), $t_i,t'_i,t''_i$ are distinct. Then $t_1$ is black if and only if $t_2$ is white, $t'_1$ is black if and only if $t'_2$ is white, and
$t''_1$ is black if and only if $t''_2$ is white. Without loss of generality, assume that $t_1$ is black, and $t_2$ is white. Then $t'_1$ is white, and $t'_2$ is black, but now, $t''_1$ and $t''_2$ are white, which is a contradiction.

\noindent
$(ii)$: Let $t_1t_2 \in E$, $t'_1t'_2 \in E$, be two such edges between $T_1$ and $T_2$.
By (A2), $t_1 \neq t'_1$, and $t_2 \neq t'_2$.
Then again, $t_1$ or $t'_1$ is black as well as $t_2$ or $t'_2$ is black, and thus, every other vertex in $T_1$ or $T_2$ is white.

\medskip

Thus Lemma \ref{S115fr3edgesbetweenTiTj} is shown.
\qed

\medskip

By Lemma \ref{S115fr3edgesbetweenTiTj} $(i)$, we can assume:
\begin{itemize}
\item[(A5)] For $i,j \in \{1,\ldots,k\}$, $i \neq j$, there are at most two edges between $T_i$ and $T_j$.
\end{itemize}

Recall that $|T_i| \ge 2$. If there is an edge in $T_i$, say $ab \in E$ with $a,b \in T_i$ and there is a third vertex $c \in T_i$ then either $a$ or $b$ is black, and thus, by Lemma \ref{lemm:structure2} $(i)$, $c$ is forced to be white, and by the Vertex Reduction and by Lemma \ref{S115fr3edgesbetweenTiTj} $(ii)$, we can assume:

\begin{itemize}
\item[(A6)] If there is an edge in $T_i$ then $|T_i|=2$. Analogously, if there are two edges between $T_i$ and $T_j$ then $|T_i|=2$ and $|T_j|=2$.
\end{itemize}

Then let us introduce the following forcing rules (which are correct).
Since no edge in $N_3$ is in $M$ (recall (\ref{noMedgesN3N4})), we have:
\begin{itemize}
\item[(R1)] All $N_3$-neighbors of a black vertex in $N_3$ must be colored white, and all $N_3$-neighbors of a white vertex in
$N_3$ must be colored black.
\end{itemize}

Moreover, we have:
\begin{itemize}
\item[(R2)] Every $T_i$, $i \in \{1,\ldots,k\}$, should contain exactly one vertex which is black. Thus, if $t \in T_i$ is black then all the remaining vertices in $T_i \setminus \{t\}$ must be colored white.

\item[(R3)] If all but one vertices of $T_i$, $1 \le i \le k$, are white and the final vertex $t \in T_i$ is not yet colored, then $t$ must be colored black.
\end{itemize}

Since no edge between $N_3$ and $N_4$ is in $M$ (recall (\ref{noMedgesN3N4})), we have:
\begin{itemize}
\item[(R4)] For every edge $st \in E$ with $t \in N_3$ and $s \in N_4$, $s$ is white if and only if $t$ is black and vice versa.
\end{itemize}

Subsequently, for checking if $G$ has a d.i.m.\ $M$ with $xy \in M$, we consider the cases $N_4 = \emptyset$ and $N_4 \neq \emptyset$.

\medskip

Then let us introduce the following recursive algorithm which formalizes the approach we will adopt to check if $G$ has a d.i.m.\\

\noindent
{\bf Algorithm DIM($G$)}

\noindent
{\bf Input.} A connected $P_9$- and ($K_4$,diamond,butterfly)-free graph $G = (V,E)$.

\noindent
{\bf Output.} A d.i.m.\ of $G$ or the proof that $G$ has no d.i.m.

\begin{itemize}
\item[(A)] Compute a central vertex, say $x$, of $G$ such that $dist_G(x,u) \leq 4$ for every $u \in V$ and every edge $xy \in E$ is part of a $P_3$ of $G$.

\item[(B)] For each edge $xy \in E$ of $G$ [contained in a $P_3$ of $G$] do:
\begin{itemize}
\item[(B.1)] compute the distance levels $N_i$ with respect to $xy$ and apply the reduction steps as shown above: if no contradiction arose and if assumptions (A1)-(A6) hold, then go to Step (B.2), else take another edge with $x$;

\item[(B.2)] check if $G$ has a d.i.m.\ $M$ with $xy \in M$; if $yes$, then return it, and STOP.
\end{itemize}
\item [(C)] Apply the Vertex Reduction to $x$ [and in particular remove $x$]; let $G'$ denote the resulting graph, where the neighbors of $x$ in $G$ are colored by black; if $G'$ is disconnected, then execute Algorithm DIM($H$) for each connected component $H$ of $G'$; otherwise, go to Step (B), with $G := G'$.

\item [(D)] Return ``$G$ has no d.i.m.'' and STOP.  \qed
\end{itemize}

\medskip

Then, by the above, Algorithm DIM($G$) is correct and can be executed in polynomial time as soon as Step (B.2) can be so.

\medskip

Then in what follows let us try to show that Step (B.2) can be solved in polynomial time, with the agreement that $G$ is ($K_4$,diamond,butterfly)-free and enjoys  assumptions (A1)-(A6): in particular recall (\ref{N5empty}) that $N_k = \emptyset$ for $k \geq 5$.

Thus we consider the cases $N_4 = \emptyset$ and $N_4 \neq \emptyset$. Let $A_{xy} := \{x,y\} \cup N_1 \cup N_2 \cup N_3$, and recall $N_4=V \setminus A_{xy}$.

\section{The Case $N_4 = \emptyset$}\label{N4empty}

In this section, we show that for the case $N_4=\emptyset$, one can check in polynomial time whether $G$ has a d.i.m.\ $M$ with $xy \in M$;
we consider the feasible $xy$-colorings for $G[A_{xy}]$.
Recall that for every edge $uv \in M$, $u$ and $v$ are black, for $I=V(G) \setminus V(M)$, every vertex in $I$ is white, $N_2 = S_2 = \{u_1,\ldots,u_k\}$ and all $u_i$, $1 \le i \le k$, are black, $T_i=N(u_i) \cap N_3$, and recall assumptions (A1)-(A6) and rules (R1)-(R4). In particular, by (A1), $S_3 = \emptyset$, i.e., $N_3=T_1 \cup \ldots \cup T_k$.

\medskip

Clearly, in the case $N_4 = \emptyset$, all the components of $G[S_2 \cup N_3]$ can be independently colored.
Every component with at most three $S_2$-vertices has a polynomial number of feasible $xy$-colorings. Thus, we can focus on components $K$ with at least four $S_2$-vertices.

\medskip

A $P_2$ $(u,v)$ in $G[N_3]$ is {\em isolated} in $G[N_3]$ if it is not part of a $P_3$ in $G[N_3]$.

\begin{clai}\label{onlyisolP2inK}
If every $P_2$ in component $K$ in $G[S_2 \cup N_3]$ is isolated then $K$ has at most three $S_2$-vertices.
\end{clai}

\noindent
{\bf Proof.}
Suppose to the contrary that $K$ has at least four $S_2$-vertices, say $u_1,u_2,u_3,u_4$, and without loss of generality, assume that $T_2$ contacts $T_1$ and $T_3$, say $t'_1t_2 \in E$ and $t'_2t_3 \in E$ for $t'_1 \in T_1$, $t_2,t'_2 \in T_2$, and $t_3 \in T_3$. By the isolated edges, $(u_1,t'_1,t_2,u_2,t'_2,t_3,u_3)$ induce a $P_7$.

\noindent
{\bf Case 1.} $T_4$ contacts $T_1$ or $T_3$.

Without loss of generality, assume that $T_4$ contacts $T_3$, i.e., there is a $t_4 \in T_4$ which contacts a vertex in $T_3$. Clearly, $t_3t_4 \notin E$ since $t'_2t_3 \in E$ is isolated. Then $t'_3t_4 \notin E$ for a second vertex $t'_3 \in T_3$, and clearly, $t'_3$ and $t_4$ do not contact the edges $t'_1t_2$ and
$t'_2t_3$ but then $(u_1,t'_1,t_2,u_2,t'_2,t_3,u_3,t'_3,t_4)$ induce a $P_9$, which is a contradiction.

\medskip

\noindent
{\bf Case 2.} $T_4$ contacts $T_2$ but does not contact $T_1$ and $T_3$.

Let $t_4 \in T_4$ contact $T_2$. Clearly, by the isolated edges, $t_4$ does not contact $t_2,t'_2 \in T_2$. Thus assume that $t''_2t_4 \in E$ for a third vertex
$t''_2 \in T_2$. By (A3), there is a second vertex $t'_4 \in T_4$ and a second vertex $t'_3 \in T_3$, and clearly, $t_3t'_3 \notin E$, $t_4t'_4 \notin E$ and $t'_4$ does not contact $T_3$ and $t'_4$ does not contact $t'_2,t''_2 \in T_2$. But then $(t'_4,u_4,t_4,t''_2,u_2,t'_2,t_3,u_3,t'_3)$ induce a $P_9$, which is a contradiction.

\medskip

Thus, Claim \ref{onlyisolP2inK} is shown.
\qed

\medskip

From now on, we can assume that there is at least one $P_3$ with contact between $T_i$ and $T_{i+1}$ in $K$.

\begin{clai}\label{2P3contactinK}
For any $P_3$'s $(a,b,c)$ and $(d,e,f)$ in $G[N_3]$ such that $d,e,f$ are not in the $T_i$'s of $a,b,c$, there is an edge between $\{a,b,c\}$ and $\{d,e,f\}$.
\end{clai}

\noindent
{\bf Proof.}
Suppose to the contrary that there is no such edge between the $P_3$'s $(a,b,c)$ and $(d,e,f)$ in $G[N_3]$. From Lemma \ref{lemm:structure2} $(ii)$,
$a,b,c$ are in at least two $T_i$'s; assume that $a \in T_1$. Then, by Lemma \ref{lemm:structure2} $(v)$ and since $ab \in E$, $bc \in E$, we have $c \notin T_1$; let $c \in T_2$, i.e., $u_1c \notin E$. Then either $b \notin T_1$ or $b \notin T_2$; without loss of generality, let $b \notin T_1$. Analogously, since $d,e,f$ are not in $T_1 \cup T_2$, assume that $d \in T_3$ and $e,f \notin T_3$.

Let $P$ be any induced path in $G$ between $u_1$ and $u_3$ through $N_1 \cup \{x,y\}$. Then the subgraph of $G$ induced by $(c,b,a,u_1)$, $P$, and $(u_3,d,e,f)$ contains an induced $P_9$, which is a contradiction. Thus, Claim \ref{2P3contactinK} is shown.
\qed

\medskip

For a $P_5$ $P=(a,b,c,d,e)$ in $G[N_3]$ with $a \in T_i$ and $b,c,d,e \notin T_i$, vertex $a$ is a {\em special $P_5$-endpoint} of $P$ in $G[N_3]$.

\begin{clai}\label{N3noP5endvTi}
There is no $P_5$ $(a,b,c,d,e)$ in $G[N_3]$ with special $P_5$-endpoint $a$.
\end{clai}

\noindent
{\bf Proof.} Suppose to the contrary that $(a,b,c,d,e)$ is a $P_5$ in $G[N_3]$ with special $P_5$-endpoint $a \in T_i$ and $b,c,d,e \notin T_i$.
But then by Observation \ref{N4P6}, vertex $a$ is the midpoint of a $P_9$, which is a contradiction. Thus, Claim \ref{N3noP5endvTi} is shown.
\qed

\begin{clai}\label{C6inS2N3}
If $C=(t_i,u_i,t'_i,t_j,t_h,t_{\ell})$ is a $C_6$ in $G[S_2 \cup N_3]$ with exactly one vertex $u_i \in S_2$ and $t_i,t'_i \in T_i$, $t_j \in T_j$, $t_h \in T_h$, $t_{\ell} \in T_{\ell}$ (possibly $j=h$ or $h = {\ell}$) then $t_j$ and $t_{\ell}$ are $xy$-forced to be black, i.e., $u_jt_j$ and $u_{\ell}t_{\ell}$ are $xy$-forced $M$-edges, and thus, $T_j$ and $T_{\ell}$ are completely colored.
\end{clai}

\noindent
{\bf Proof.} By (\ref{noMedgesN3N4}), no edge in $N_3$ is in $M$. By Observation \ref{dimC3C5C7C4} $(iii)$, either exactly two or none of the edges in $C$
are in $M$. Since $C$ has exactly one vertex $u_i \in S_2$, $u_it_i$ and $u_it'_i$ are the only edges of $C$ which are not in $N_3$, and clearly, either $u_it_i \notin M$ or $u_it'_i \notin M$. Thus, by Observation \ref{dimC3C5C7C4} $(iii)$, no edge in $C$ is in $M$, i.e., $t_i$ and $t'_i$ are white, and $t_j$ as well as $t_{\ell}$ are $xy$-forced to be black.
Thus, Claim \ref{C6inS2N3} is shown.
\qed

\medskip

After the Edge Reduction step, we can assume that there is no such $C_6$ in $G[S_2 \cup N_3]$, i.e., every $C_6$ in $G[S_2 \cup N_3]$ has either two vertices of $S_2$ or none of it.

\begin{clai}\label{C7inS2N3}
If $C$ is a $C_7$ in $G[S_2 \cup N_3]$ then $C$ has exactly two vertices in $S_2$, say $C=(t_i,u_i,t'_i,t_j,u_j,t'_j,t_h)$, and then
$t_h$ is $xy$-forced to be black, i.e., $u_ht_h$ is an $xy$-forced $M$-edge.
\end{clai}

\noindent
{\bf Proof.}
Let $C$ be a $C_7$ in $G[S_2 \cup N_3]$. Recall that by Lemma \ref{lemm:structure2} $(iii)$, there is no $C_7$ in $G[N_3]$. Thus, $|V(C) \cap S_2| \ge 1$, and clearly, by (A1), no vertex in $V(C) \cap N_3$ contacts two vertices in $V(C) \cap S_2$, i.e., $|V(C) \cap S_2| \le 2$.

If there is exactly one $S_2$-vertex in a $C_7$ in $G[N_3]$, say $C=(t_1,u_1,t'_1,t_2,t',t'',t''')$ with $t_1,t'_1 \in T_1$ then $t_2,t',t'',t''' \notin T_1$, but now, $(t'_1,t_2,t',t'',t''')$ induce a $P_5$ with special $P_5$-endpoint $t'_1 \in T_1$ such that $t_2,t',t'',t''' \notin T_1$, which is a contradiction to Claim \ref{N3noP5endvTi}.

Now assume that $C=(t_1,u_1,t'_1,t_2,u_2,t'_2,t_3)$ is a $C_7$ in $G[S_2 \cup N_3]$. Suppose to the contrary that $t_3$ is white. Then $t_1$ and $t'_2$ are black which implies that $t'_1$ and $t_2$ are white, which is a contradiction since $t'_1t_2 \in E$. Thus, $t_3$ is $xy$-forced to be black, i.e., $u_3t_3$ is an $xy$-forced $M$-edge, and Claim \ref{C7inS2N3} is shown.
\qed

\medskip

After the Edge Reduction step, we can assume that there is no $C_7$ in $G[S_2 \cup N_3]$.

\begin{clai}\label{C9inS2N3}
If there is a $C_9$ $C$ in $G[S_2 \cup N_3]$ then $|V(C) \cap S_2|=3$, say $V(C) \cap S_2=\{u_1,u_2,u_3\}$, and for the component $K$ in $G[S_2 \cup N_3]$ containing $C$, we have\\
$K=G[\{u_1,u_2,u_3\} \cup T_1 \cup T_2 \cup T_3]$.
\end{clai}

\noindent
{\bf Proof.}
Let $C$ be a $C_9$ in $G[S_2 \cup N_3]$. Recall that by Lemma \ref{lemm:structure2} $(iii)$, there is no $C_9$ in $G[N_3]$, i.e., $|V(C) \cap S_2| \ge 1$, and clearly, $|V(C) \cap S_2| \le 3$.

If $C$ contains only one $S_2$-vertex then, as in the proof of Claim \ref{C7inS2N3}, it leads to a $P_5$ in $N_3$ with corresponding special $P_5$-endpoint, which is a contradiction to Claim \ref{N3noP5endvTi}. Thus, $|V(C) \cap S_2| \ge 2$.

First assume that $|V(C) \cap S_2|=2$. If $C=(t_1,u_1,t'_1,t_2,u_2,t'_2,t_3,t_4,t_5)$ (possibly $t_3,t_4 \in T_3$ or $t_4,t_5 \in T_4$) then this leads to a $P_5$ $(t'_2,t_3,t_4,t_5,t_1)$ with special $P_5$-endpoint $t_1$, which is a contradiction to Claim \ref{N3noP5endvTi}.
If $C=(t_1,u_1,t'_1,t_2,t_3,u_3,t'_3,t_4,t_5)$ then $t_2$ is $xy$-forced to be black: Suppose to the contrary that $t_2$ is white. Then $t'_1$ and $t_3$ are black, which implies that $t_1$ and $t'_3$ are white, but now, $t_4$ and $t_5$ are black, which is a contradiction since there is no $M$-edge in $N_3$.
Thus, $u_2t_2$ is an $xy$-forced $M$-edge, and after the Edge Reduction, $|V(C) \cap S_2|=2$ is impossible.

Thus, $|V(C) \cap N_2|=3$; let $C=(t_1,u_1,t'_1,t_2,u_2,t'_2,t_3,u_3,t'_3)$ be a $C_9$ with three such $S_2$-vertices $u_1,u_2,u_3$.
Suppose to the contrary that there is a vertex $t_4 \in T_4$ which contacts $C$, say $t'_3t_4 \in E$.
Clearly, by Lemma \ref{lemm:structure2} $(v)$, $t_4t_3 \notin E$.
Since $(u_1,t'_1,t_2,u_2,t'_2,t_3,u_3,t'_3,t_4)$ do not induce a $P_9$, we have $t_4t'_1 \in E$ or $t_4t_2 \in E$ or $t_4t'_2 \in E$.

If $t_4t'_1 \in E$ then $(t_2,t'_1,t_4,t'_3,t_1)$ would induce a $P_5$ in $N_3$ with special $P_5$-endpoint $t_2$, which is impossible by Claim~\ref{N3noP5endvTi}.
Similarly, if $t_4t'_2 \in E$ then $(t_3,t'_2,t_4,t'_3,t_1)$ would induce a $P_5$ in $N_3$ with special $P_5$-endpoint $t_1$, which is a contradiction to Claim~\ref{N3noP5endvTi}.

Thus, $t_4t_2 \in E$ which leads to a $C_7$ $(t_2,u_2,t'_2,t_3,u_3,t'_3,t_4)$. But then, by Claim \ref{C7inS2N3}, $t_4$ is $xy$-forced to be black, i.e.,
$u_4t_4$ is an $xy$-forced $M$-edge, and after the Edge Reduction, there is no such $C_7$.
Thus, Claim \ref{C9inS2N3} is shown.
\qed

\begin{corollary}\label{componwithfourS2vertC9fr}
Every component in $G[S_2 \cup N_3]$ with at least four $S_2$-vertices is $C_9$-free.
\end{corollary}

\begin{lemma}\label{P9frN4emptycoloringpol}
In the case $N_4=\emptyset$, for every component $K$ in $G[S_2 \cup N_3]$, a complete coloring of $K$ (if there is no contradiction) can be done in polynomial time.
\end{lemma}

\noindent
{\bf Proof.}
For finding a complete feasible $xy$-coloring of component $K$ (or a contradiction), we first use Vertex Reduction and Edge Reduction as in the previous results.

Let $V(K) \cap N_3 = T_1 \cup \ldots \cup T_h$ (recall that $h \ge 4$ since otherwise, a complete feasible $xy$-coloring of $K$ can be done in polynomial time). Clearly, for every $i$, $1 \le i \le h$, we have $|T_i| \ge 2$.

If every $T_i$ in $K$ would have only one out-vertex $t_i \in T_i$, then the procedure starts by fixing a coloring of $T_1$; for every $i$, $1 \le i \le h$, there are only two possible colorings of $T_i$ since by (A4), every $T_i$ has at most one in-vertex. If the already colored out-vertex $t_i \in T_i$ with contact to $t_{i+1} \in T_{i+1}$ is white then $t_{i+1}$ is black, the in-vertex of $T_{i+1}$ is white, and $T_{i+1}$ is completely colored. Analogously, if $t_i$ is black then $t_{i+1}$ is white, the in-vertex of $T_{i+1}$ is black, and $T_{i+1}$ is completely colored.

\medskip

Thus, we can assume that there is a $T_i$ with at least two out-vertices (such that at least one of them is white).
In this case, the procedure starts by fixing a coloring of $T_1$ with at least two out-vertices (this can be repeated for all $|T_1|$ colorings of $T_1$) and applies the forcing rules, and then the next step of the procedure is using a white out-vertex $t_1 \in T_1$, say with contact to $T_2$, such that the neighbor $t_2 \in T_2$ of $t_1$ is black. If $t_2$ contacts only $T_1$ then $t_2$ does not play any role for the procedure. If $t_2$ contacts some $T_j$, $j \neq 1,2$, the problem is how $T_j$ can be completely colored.

\medskip

Now we can assume that every black out-vertex (which was already colored by a white neighbor in the previous step) contacts at least two $T_i$'s, say, $t_2 \in T_2$ was colored black by a white vertex $t'_1 \in T_1$ with $t'_1t_2 \in E$ (i.e., $T_1$ was already colored), and $t_2t_3 \in E$ for $t_3 \in T_3$ but $T_3$ is not yet completely colored. Then $t_3$ is white, and if $t_3t'_3 \in E$ for another $t'_3 \in T_3$ then $t'_3$ is black and $T_3$ is completely colored. Thus assume that $t_3t'_3 \notin E$ for any $t'_3 \in T_3$.
If $t_3$ is the only out-vertex in $T_3$ then the in-vertex is black (recall that by (A4), every $T_i$ has at most one in-vertex) and $T_3$ is completely colored. Thus assume that $t'_3$ is an out-vertex, say $t'_3t_4 \in E$ for $t_4 \in T_4$.
We first show:

\begin{clai}\label{P2P3contactT3complcol}
If there is any contact between the $P_3$ $(t'_1,t_2,t_3)$ and the $P_2$ $(t'_3,t_4)$ then $T_3$ is completely colored.
\end{clai}

\noindent
{\em Proof.} Clearly, $t_3t'_3 \notin E$, $t_3t_4 \notin E$, and $t_2t'_3 \notin E$.
If $t'_1t'_3 \in E$ then $t'_3$ is black and $T_3$ is completely colored. Thus assume that $t'_1t'_3 \notin E$.
If $t_2t_4 \in E$ then $t_4$ is white and thus, $t'_3$ is black and $T_3$ is completely colored. Thus assume that $t_2t_4 \notin E$.

Finally, if $t'_1t_4 \in E$ then $(t'_1,t_2,t_3,u_3,t'_3,t_4)$ induce a $C_6$, which is impossible by Claim~\ref{C6inS2N3} and the Edge Reduction.

Thus, $T_3$ is completely colored.
\qed

\medskip

Now we assume:
$$(t'_1,t_2,t_3) \cojoin (t'_3,t_4).$$

Moreover, $(u_1,t'_1,t_2,t_3,u_3,t'_3,t_4,u_4)$ induce a $P_8$ in $G$.
Clearly, $|T_1| \ge 2$ and $|T_4| \ge 2$; let $t_1 \in T_1$ be a second vertex in $T_1$ and $t'_4 \in T_4$ be a second vertex in $T_4$.

\begin{clai}\label{t4nonadjt'4T3complcol}
If $t_4t'_4 \notin E$ then $T_3$ is completely colored.
\end{clai}

\noindent
{\em Proof.}
If $t_4t'_4 \notin E$ then clearly, $t'_3t'_4 \notin E$. Since $(u_1,t'_1,t_2,t_3,u_3,t'_3,t_4,u_4,t'_4)$ do not induce a $P_9$ in $G$, we have
$t'_1t'_4 \in E$ or $t_2t'_4 \in E$ or $t_3t'_4 \in E$. If $t_3t'_4 \in E$ then $|T_3|=2$ (recall (A6)) and $T_3$ is completely colored. Thus assume that $t_3t'_4 \notin E$.
Now, if $t_2t'_4 \in E$ then $(t_2,t_3,u_3,t'_3,t_4,u_4,t'_4)$ induce a $C_7$, which is impossible by Claim \ref{C7inS2N3} and the Edge Reduction.
Thus, $t_2t'_4 \notin E$ which implies that $t'_1t'_4 \in E$. But now, since $t'_1$ is white, $t'_4$ is black, $t_4$ is white, $t'_3$ is black, and $T_3$ is completely colored.
\qed

\medskip

From now on, we assume: $$t_4t'_4 \in E.$$

\noindent
{\bf Case 1.} $t_1t'_1 \in E$.

\medskip

Then $t_1$ is black.
By Claim \ref{2P3contactinK}, $(t_2,t'_1,t_1)$ and $(t'_3,t_4,t'_4)$ do not induce a $2P_3$. Recall that $(t_2,t'_1)$ and $(t'_3,t_4)$ induce a $2P_2$.
Thus, $t_1$ should contact $(t'_3,t_4,t'_4)$ or $t'_4$ should contact $(t_2,t'_1,t_1)$.

If $t'_1t'_4 \in E$ then $t'_4$ is black, $t_4$ is white, $t'_3$ is black, and $T_3$ is completely colored. Thus assume $t'_1t'_4 \notin E$.
Analogously, if $t_1t_4 \in E$ then $t_4$ is white, and $t'_3$ is black, and $T_3$ is completely colored. Thus assume $t_1t_4 \notin E$.

If $t_1t_3 \in E$ then clearly, $t_1t'_3 \notin E$, and if $t_1t_3 \notin E$ but $t_1t'_3 \in E$ then $(t_1,t'_1,t_2,t_3,u_3,t'_3)$ induce a $C_6$ which is impossible by Claim \ref{C6inS2N3} and the Edge Reduction. Thus, assume that $t_1t'_3 \notin E$.

If $t_2t'_4 \in E$ then $(t_2,t_3,u_3,t'_3,t_4,t'_4)$ induce a $C_6$ which is impossible by Claim \ref{C6inS2N3} and the Edge Reduction.
Thus, assume that $t_2t'_4 \notin E$.

Now, $t_1t'_4 \in E$ is the only possible edge between $(t_2,t'_1,t_1)$ and $(t'_3,t_4,t'_4)$ but now, $(t_2,t'_1,t_1,t'_4,t_4)$ induce a $P_5$ with special endpoint
 $t_2$, which is impossible by Claim \ref{N3noP5endvTi}.

Thus, in Case 1, $T_3$ is completely colored.

\medskip

\noindent
{\bf Case 2.} $t_1t'_1 \notin E$.

\medskip

Recall that $t'_1$ and $t_3$ are white and $(t'_1,t_2,t_3) \cojoin (t'_3,t_4)$. Clearly, $t_1t_2 \notin E$, and since $(t_1,u_1,t'_1,t_2,t_3,u_3,t'_3,t_4,u_4)$ do not induce a $P_9$ in $G$, we have
$t_1t_3 \in E$ or $t_1t'_3 \in E$ or $t_1t_4 \in E$.

If $t_1t'_3 \in E$ then clearly, $t_1t_3 \notin E$. But then $(t_1,u_1,t'_1,t_2,t_3,u_3,t'_3)$ induce a $C_7$, which is impossible by Claim \ref{C7inS2N3} and the Edge Reduction. Thus, assume $t_1t'_3 \notin E$ and either $t_1t_3 \in E$ or $t_1t_4 \in E$.

\medskip

\noindent
{\bf Case 2.1} $t_1t_3 \notin E$.

\medskip

Then $t_1t_4 \in E$. First assume that $t_1$ is white, which implies that $t_4$ is black, and there is a black vertex $t''_1 \in T_1$.
Clearly, $t''_1t_1 \notin E$, $t''_1t'_1 \notin E$, $t''_1t_2 \notin E$, and since $t_4$ is black, we have $t''_1t_4 \notin E$.
Since $(t''_1,u_1,t'_1,t_2,t_3,u_3,t'_3,t_4,u_4)$ do not induce a $P_9$ in $G$, we have $t''_1t_3 \in E$ or $t''_1t'_3 \in E$.

If $t''_1t'_3 \in E$ then $t''_1t_3 \notin E$, but now $(t''_1,u_1,t'_1,t_2,t_3,u_3,t'_3)$ induce a $C_7$, which is impossible by Claim \ref{C7inS2N3} and the Edge Reduction. Thus, assume $t''_1t'_3 \notin E$ which implies $t''_1t_3 \in E$. But now $(t''_1,u_1,t_1,t_4,t'_3,u_3,t_3)$ induce a $C_7$, which is impossible by Claim \ref{C7inS2N3} and the Edge Reduction.

\medskip

Thus $t_1$ is black which implies that $t_4$ is white, $t'_3$ is black, $T_3$ is completely colored, and Case 2.1 is done.

\medskip

\noindent
{\bf Case 2.2} $t_1t_3 \in E$.

\medskip

Since $t_3$ is white, $t_1$ is black. Clearly, since $t_1t_3 \in E$, we have $t_1t'_3 \notin E$. If $t_1t_4 \in E$ then $t_4$ is white and thus, $t'_3$ is black and $T_3$ is completely colored. Thus, assume that $t_1t_4 \notin E$.

\medskip

If $|T_1| \ge 3$ then let $t''_1 \in T_1$ be a second white vertex in $T_1$.
Then $t''_1t_3 \notin E$, and Case~2.1 applies for $t''_1 \in T_1$.
Thus we have $$|T_1| = 2.$$

Next assume that the black out-vertex $t_1$ has a second white neighbor, say $t_0 \in T_0$ with $t_0t_1 \in E$.
Since $(t_0,t_1,t_3,t_2,t'_1)$ do not induce a $P_5$ with special endpoint $t_0$ (recall Claim \ref{N3noP5endvTi}), we have $t_0t_2 \in E$.
Recall that $t'_1t'_4 \notin E$ (else  $T_3$ is completely colored) and $t_2t'_4 \notin E$ (else $(t_2,t_3,u_3,t'_3,t_4,t'_4)$ induce a $C_6$ which is impossible by Claim \ref{C6inS2N3} and the Edge Reduction).

Since $(t'_1,t_2,t_0)$ and $(t'_3,t_4,t'_4)$ do not induce a $2P_3$ (recall Claim \ref{2P3contactinK}), we have $t_0t'_3 \in E$ or $t_0t_4 \in E$ or $t_0t'_4 \in E$.
If $t_0t'_3 \in E$ or $t_0t'_4 \in E$ then $t'_3$ is black and $T_3$ is completely colored. Thus assume that $t_0t_4 \in E$.
But now, $(t_0,t_1,t_3,u_3,t'_3,t_4)$ induce a $C_6$ which is impossible by Claim \ref{C6inS2N3} and the Edge Reduction.
Thus, $t_1$ has only one white neighbor, namely $t_3$.

\medskip

\noindent
Next we show:
\begin{clai}\label{t'4nooutvertex}
$t'_4$ is no out-vertex.
\end{clai}

\noindent
{\em Proof.}
Suppose to the contrary that $t'_4t_5 \in E$ for some $t_5 \in T_5$. Recall that $t_4$ does not contact $t'_1,t_2,t_3$.
Since $t'_4$ is white (else $T_3$ is completely colored), $t'_4$ does not contact $t'_1,t_3$, and recall that $t'_4t_2 \notin E$
(else $(t_2,t_3,u_3,t'_3,t_4,t'_4)$ induce a $C_6$, which is impossible by Claim~\ref{C6inS2N3} and the Edge Reduction).

Recall that $(t'_1,t_2,t_3)$ and $(t_4,t'_4,t_5)$ do not induce a $2P_3$. Thus, since $t_4$ and $t'_4$ do not contact $(t'_1,t_2,t_3)$,
only $t_5$ could contact $(t'_1,t_2,t_3)$. Since $t_2$ and $t_5$ are black, we have $t_5t_2 \notin E$.
If $t_5t_3 \in E$ then clearly, $t_5t'_3 \notin E$ but then $(t_3,u_3,t'_3,t_4,t'_4,t_5)$ induce a $C_6$, which is impossible by Claim \ref{C6inS2N3} and the Edge Reduction. Thus, $t_5t_3 \notin E$.
Now, if $t_5t'_1 \in E$ then $(t_1,t_3,t_2,t'_1,t_5)$ induce a $P_5$ with special endpoint $t_5$, which is a contradiction by Claim \ref{N3noP5endvTi}.

Thus $t_5t'_1 \notin E$, $t_5t_2 \notin E$, and $t_5t_3 \notin E$. But then $(t'_1,t_2,t_3)$ and $(t_4,t'_4,t_5)$ induce a $2P_3$, which is a contradiction by Claim \ref{2P3contactinK}. Thus, Claim \ref{t'4nooutvertex} is shown.
\qed

\medskip

This implies that $t_4$ is the only out-vertex in $T_4$.

\medskip

\noindent
Next we show:
\begin{clai}\label{t2contactsonlyoneTi}
The black vertex $t_2$ contacts only one $T_i$ (namely $T_3$) which is not yet completely colored.
\end{clai}

\noindent
{\em Proof.}
Suppose to the contrary that $t_2$ contacts a second $T_i$ (apart from $T_3$) which is not yet completely colored. Then $T_i \neq T_4$ since by the above $T_4 = \{t_4,t'_4\}$ and $t_2$ is nonadjacent to $t_4,t'_4$. Then say $T_i = T_5$, i.e., $t_2$ contacts $T_5$ with $t_2t_5 \in E$. Note that $t_5$ does not contact $T_3$ (else $T_3$ is completely colored) and does not contact $T_4$ (else, by Claim \ref{t'4nooutvertex}, $t_5t_4 \in E$ and then vertices $t_2,t_3,u_3,t'_3,t_4,t_5$ would induce a $C_6$ according to Claim \ref{C6inS2N3}). Since $T_5$ is not yet completely colored, $T_5$ is not contacted by $t'_1$ and $t_3$, and furthermore (similarly to the above with respect to $T_3$) there is an out-vertex of $T_5$ say $t'_5 \in T_5$ (nonadjacent to $t_5$).
Note that $t'_5$ does not contact $T_3$ (else, say $t'_5t''_3 \in E$ with $t''_3 \in T_3$, vertices $t_2,t_3,u_3,t''_3,t'_5,u_5,t_5$ induce a $C_7$ in $G[S_2 \cup N_3]$) and does not contact $T_4$ (else, by Claim \ref{t'4nooutvertex}, $t'_5t_4 \in E$ and then vertices $u_1,t'_1,t_2,t_5,u_5,t'_5,t_4,t'_3,u_3$ induce a $P_9$). Then vertices $t'_5,u_5,t_5,t_2,t_3,u_3,t'_3,t_4,t'_4$ induce a $P_9$, 
which is a contradiction.

Thus, Claim~\ref{t2contactsonlyoneTi} is shown.
\qed

\medskip

\noindent
Finally we show:
\begin{clai}\label{onlyoneblackcontactsTi}
There is only one black vertex which contacts a $T_i$ which is not yet completely colored.
\end{clai}

\noindent
{\em Proof.}
Suppose to the contrary that there are two such black vertices, say $t_2$ which contacts $T_3$ and $t_6$ which contacts $T_7$ such that $T_3,T_7$ are not yet completely colored.
By Claim~\ref{t2contactsonlyoneTi}, $t_2$ does not contact $T_7$ and $t_6$ does not contact $T_3$. If $t'_1 \neq t'_5$ then clearly, the $P_3$'s $(t'_1,t_2,t_3)$ and $(t'_5,t_6,t_7)$ do not induce a $2P_3$ but $t_2t_7 \notin E$ and $t_6t_3 \notin E$. Thus, $t'_1t_6 \in E$ or $t'_5t_2 \in E$, say without loss of generality, $t'_1t_6 \in E$ but now, $(t_3,t_2,t'_1,t_6,t_7)$ induce a $P_5$ with special endpoint $t_3$, which is a contradiction to Claim \ref{N3noP5endvTi}. Analogously, if $t'_1=t'_5$, it leads to the same contradiction.
Thus, Claim \ref{onlyoneblackcontactsTi} is shown.
\qed

\medskip

In general, if $t_2$ does not completely color $T_3$ then we can add a possible coloring of $T_3$ which leads to a complete coloring of every neighbor $T_i$ of $T_3$.
Since $G$ is $P_9$-free, Case~2.2 appears only once in component $K$.

\medskip

Thus, Lemma \ref{P9frN4emptycoloringpol} is shown.
\qed

\section{The Case $N_4 \neq \emptyset$}\label{N4nonempty}

Recall $A_{xy} := \{x,y\} \cup N_1 \cup N_2 \cup N_3$ and $N_5 = \emptyset$. In the case $N_4 \neq \emptyset$, we show that one can check in polynomial time whether $G$ has a d.i.m.\ $M$ with $xy \in M$. Clearly, again in the case $N_4 \neq \emptyset$, all the components of $G[S_2 \cup N_3 \cup N_4]$ can be independently colored.

Recall Observation \ref{N4P6}; if $t \in N_3$ then $t$ is an endpoint of a corresponding induced $P_5$ in $\{x,y\} \cup N_1 \cup S_2 \cup N_3$. If there is a $P_5$ $(t,a,b,c,d)$ with endpoint $t$ and four vertices $a,b,c,d \in N_4$ (such that only one of them, say $a$ contacts $t$) then $t$ is the midpoint of a $P_9$ in $G$, which is a contradiction. Analogously, if there is a $P_5$ $(t,a,b,c,t')$ with $t,t' \in N_3$ such that $t$ and $t'$ are in distinct $T_i$'s then $t$ is the midpoint of a $P_9$ in $G$, which is a contradiction. This argument is used in some of the next proofs.

\begin{proposition}\label{prop:N4}
If the colors of all vertices in $G[A_{xy}]$ are fixed then the colors of all vertices in $N_4$ are forced.
\end{proposition}

\noindent
{\bf Proof.} Let $v \in N_4$ and let $w \in N_3$ be a neighbor of $v$. Since by (\ref{noMedgesN3N4}), every edge between $N_3$ and $N_4$ is $xy$-excluded, we have:
If $w$ is white then $v$ is black, and if $w$ is black then $v$ is white.
\qed

\medskip

Let $K$ be a nontrivial component of $G[S_2 \cup N_3 \cup N_4]$. Clearly, $K$ can have several components in $G[S_2 \cup N_3]$ which are connected by some $N_4$-vertices. $K$ can be feasibly colored (if there is no contradiction) by starting with a component in $G[S_2 \cup N_3]$ or with a component in $G[N_4]$ which is part of $K$.

\medskip

Recall (\ref{triangleaN3bcN4}) and (\ref{edgeN4N3neighb}) for the fact that after the Edge Reduction, there is no triangle between $N_3$ and $N_4$ with exactly one vertex in $N_3$, and for every edge $uv$ in $G[N_4]$, $u$ and $v$ have no common neighbor in $N_3$.
Moreover, for $N_4$-vertices which are isolated in $N_4$, we have:
\begin{equation}\label{N4isolatedvertexwhite}
\mbox{If } v \in N_4 \mbox{ with } N(v) \cap N_4=\emptyset \mbox{ then } v \mbox{ is white}.
\end{equation}

Thus, after the Vertex Reduction, we can assume that every vertex in $N_4$ has a neighbor in $N_4$, i.e., every component in $G[N_4]$ has at least one edge.

\medskip

Similarly, we have:
\begin{clai}\label{N4isolatededgeblack}
If $v,w \in N_4$ with $vw \in E$ is an $N_4$-isolated edge in $G[N_4]$ then $vw$ is an $xy$-forced $M$-edge, i.e., $v$ and $w$ are black.
\end{clai}

\noindent
{\bf Proof.} Let $v,w \in N_4$ with $vw \in E$ such that $v$ and $w$ do not have any other neighbors in $N_4$.
Clearly, since $vw \in E$, at least one of $v$ and $w$ is black, say $v$ is black. If $w$ is white then $v$ needs a black $M$-mate in $N_4$ since by
(\ref{noMedgesN3N4}), there is no $M$-edge between $N_3$ and $N_4$. But since $vw$ is $N_4$-isolated, there is no such $M$-mate of $v$, i.e., $w$ is black.
Thus, Claim~\ref{N4isolatededgeblack} is shown.
\qed

\medskip

Thus, after the Edge Reduction, there is no such $N_4$-isolated edge in $N_4$.

\medskip

By the way, there are possible contradictions: For instance, if for an $N_4$-isolated edge $vw$, $v$ or $w$ contacts a black vertex in $N_3$ then there is no d.i.m.\ with $xy \in M$.
Analogously, if $vt \in E$ for $t \in N_3$ and $wt' \in E$ for $t' \in N_3$ and $tt' \in E$ (i.e., $(t,v,w,t')$ induce a $C_4$) then there is no d.i.m.\ with $xy \in M$.

\begin{clai}\label{C5N3N4oneedgeinN4}
If $t \in T_i$, $t' \in T_j$ (possibly $i=j$), and $a,b,c \in N_4$ induce a $C_5$ $C=(t,a,b,t',c)$ in $G[N_3 \cup N_4]$ then $ab$ is an $xy$-forced $M$-edge.
\end{clai}

\noindent
{\bf Proof.}
Let $C=(t,a,b,t',c)$ be a $C_5$ in $G[N_3 \cup N_4]$. Then the edges $ta,tc,t'b,t'c$ are edges between $N_3$ and $N_4$. By Observation
\ref{dimC3C5C7C4} $(i)$, every $C_5$ has exactly one $M$-edge, and by (\ref{noMedgesN3N4}), no edge between $N_3$ and $N_4$ is in $M$. Thus, $ab$ is an $xy$-forced $M$-edge, and Claim \ref{C5N3N4oneedgeinN4} is shown.
\qed

\medskip

In general, for any $C_5$ in $G[N_3 \cup N_4]$ with exactly one edge in $G[N_4]$, this edge is $xy$-forced as an $M$-edge. After the Edge Reduction step, we can assume that there is no such $C_5$ in $G[N_3 \cup N_4]$ with exactly one edge in $G[N_4]$.

\begin{corollary}\label{P4midedgeinN4}
If $ab \in E$ for $a,b \in N_4$ and $at \in E$, $bt' \in E$ for $t,t' \in N_3$, $t \neq t'$, such that $(t,a,b,t')$ induce a $P_4$ in $G$ then there is no common neighbor $c \in N_4$ of $t$ and $t'$.
\end{corollary}

\noindent
{\bf Proof.}
Suppose to the contrary that there is such a common neighbor $c \in N_4$ with $tc \in E$ and $t'c \in E$. Then $ac \notin E$ and $bc \notin E$ since there are no triangles $(t,a,c)$, $(t',b,c)$. But then $C=(t,a,b,t',c)$ induce a $C_5$, which is a contradiction by Claim \ref{C5N3N4oneedgeinN4} and the Edge Reduction. Thus, Corollary \ref{P4midedgeinN4} is shown.
\qed

\begin{clai}\label{C4N3N4P3}
If $t \in T_i$ and $a,b,c \in N_4$ induce a $C_4$ $C=(t,a,b,c)$ then $t$ is black and $u_it$ is an $xy$-forced $M$-edge.
\end{clai}

\noindent
{\bf Proof.}
Let $C=(t,a,b,c)$ be a $C_4$ with exactly one $N_3$-vertex $t$. Suppose to the contrary that $t$ is white.
Then by Observation \ref{dimC3C5C7C4} $(ii)$, $a$ and $c$ are black, $b$ is white, and by (\ref{noMedgesN3N4}), there are $M$-mates $a' \in N_4$ of $a$ and $c' \in N_4$ of $c$, i.e., $aa' \in M$ and $cc' \in M$. Since $G$ is butterfly-free, $b$ is nonadjacent to at least one vertex of $\{a',c'\}$, say $b$ is nonadjacent to $c'$ without loss of generality by symmetry. By (\ref{triangleaN3bcN4}) and the Edge Reduction, $tc' \notin E$; let $t' \in N_3$ be a neighbor of $c'$, i.e., $t'c' \in E$.
Then $t'$ is white (since $cc' \in M$), i.e., $t'b \notin E$; furthermore, by (\ref{noMedgesN3N4}), $t'c \notin E$. Then, since $(t',c',c,b,a)$ do not induce a $P_5$ (else by Observation \ref{N4P6}, $t'$ is the midpoint of a $P_9$ in $G$), we have $t'a \in E$ but now, $C=(t,a,t',c',c)$ is a $C_5$ with exactly one edge in $N_4$, namely $cc'$. By Claim \ref{C5N3N4oneedgeinN4} and the Edge Reduction, we have that there is no such $C_5$, i.e., $t$ is black and $u_it$ is an $xy$-forced $M$-edge.
Thus, Claim \ref{C4N3N4P3} is shown.
\qed

\medskip

After the Edge Reduction step, we can assume that there is no such $C_4$ in $G[N_3 \cup N_4]$.

\begin{corollary}\label{P4inN4C5}
\mbox{ }
\begin{enumerate}
\item[$(i)$] If $(a,b,c,d)$ induce a $P_4$ in $G[N_4]$ with $N_3$-neighbor $t$ of $a$ then $(t,a,b,c,d)$ induce a $C_5$ in $G[N_3 \cup N_4]$.
\item[$(ii)$] If $(a,b,c)$ induce a $P_3$ in $G[N_4]$ with $N_3$-neighbor $t$ of $a$ and $t'$ of $c$ (clearly, $t \neq t'$) then either $tt' \in E$, i.e., $(t,a,b,c,t')$ induce a $C_5$ in $G[N_3 \cup N_4]$, or $t,t' \in T_i$.
\end{enumerate}
\end{corollary}

\begin{clai}\label{N4noP3abcwhiteac}
There is no $P_3$ $(a,b,c)$ in $G[N_4]$ with white end-vertices $a$ and $c$.
\end{clai}

\noindent
{\bf Proof.}
Suppose to the contrary that there is such a $P_3$ $(a,b,c)$ in $G[N_4]$ with white end-vertices $a$ and $c$, and thus black vertex $b$. Let $t_a \in T_i$ be an $N_3$-neighbor of $a$, and let $t_c$ be an $N_3$-neighbor of $c$. By Claim \ref{C4N3N4P3} and the Edge Reduction, $t_ac \notin E$ and $t_ca \notin E$, i.e., $t_a \neq t_c$. Clearly, $t_a$ and $t_c$ are black, and thus, $t_c \notin T_i$ (and there is no $T_j$ with $t_a,t_c \in T_j$). Moreover, $t_at_c \notin E$ since both of them are black (recall that by (\ref{noMedgesN3N4}), there is no $M$-edge in $N_3$). But then $(t_a,a,b,c,t_c)$ induce a $P_5$, and it leads to a $P_9$ in $G$ with midpoint $t_a$, which is a contradiction.
Thus, Claim~\ref{N4noP3abcwhiteac} is shown.
\qed

\begin{corollary}\label{N4degree3white}
If vertex $z$ in $G[N_4]$ has degree at least $3$ in $G[N_4]$ then $z$ is white.
\end{corollary}

\noindent
{\bf Proof.}
Suppose to the contrary that there is a black vertex $z$ in $G[N_4]$ with degree at least $3$, say $zz_i \in E$, $1 \le i \le 3$. Without loss of generality, assume that $z_1$ is black. But then $z_2$ and $z_3$ are white, and thus, $(z_2,z,z_3)$ induce a $P_3$ with white end-vertices $z_2,z_3$, which is a contradiction to Claim~\ref{N4noP3abcwhiteac}.
Thus, Corollary~\ref{N4degree3white} is shown.
\qed

\medskip

Thus, after the Vertex Reduction, every vertex in a component of $G[N_4]$ has degree at most 2 in $G[N_4]$. For every component $D$ of $G[N_4]$, this leads to  feasible colorings of $D$:

\begin{clai}\label{componinN4}
Every component $D$ in $G[N_4]$ is either a $P_k$, $3 \le k \le 8$, or a $C_k$, $k \in \{3,6,9\}$, and $D$ has at most three feasible colorings.
\end{clai}

\noindent
{\bf Proof.} Recall that after the Vertex Reduction, every vertex in a component $D$ of $G[N_4]$ has degree at most 2 in $G[N_4]$. If $D$ is cycle-free then, since $D$ contains a $P_3$ (recall (\ref{N4isolatedvertexwhite}) and Claim \ref{N4isolatededgeblack}) and $G$ is $P_9$-free, $D$ is a $P_k$, $3 \le k \le 8$. If $D$ contains a $C_k$ $C$ then, since $G$ is $P_9$-free, $k \le 9$, and since every vertex in $C$ has degree 2, $C$ is no $C_4$, $C_5$, $C_7$, $C_8$, since every black vertex in $C$ must have an $M$-mate in $C$. Thus, $C$ is either a
$C_3$, $C_6$, or $C_9$.
Clearly, for a $C_k$, $k \in \{3,6,9\}$, say $C=(z_1,\ldots,z_k)$, there are three feasible colorings; for example, in a $C_9$, if $z_1$ is white then $z_4$ and $z_7$ are white and the remaining vertices are black, and similarly if $z_2$ is white or $z_3$ is white. For induced paths $P_k$, $3 \le k \le 8$, say $P=(z_1,\ldots,z_k)$, there are either one or two feasible colorings; if $z_1$ is white then $z_2$ and $z_3$ are black and thus $z_4$ is white etc. Thus, it leads to exactly one feasible coloring for $P_4$, $P_5$, $P_7$, $P_8$, and for exactly two feasible colorings for $P_3$ and $P_6$.
Thus, Claim~\ref{componinN4} is shown.
\qed

\begin{clai}\label{N4P3abcwhiteb}
Let $(a,b,c)$ be a $P_3$ in $G[N_4]$ for $a,b,c \in N_4$.
If $b$ is white then all $N_3$-neighbors of $a,b,c$ are in the same $T_i$.
\end{clai}

\noindent
{\bf Proof.}
Let $t_a \in N_3$ be the neighbor of $a$, and analogously, let $t_b$, $t_c$ be the neighbors of $b$, $c$ in $N_3$. Without loss of generality, let $t_a \in T_1$.
Clearly, $t_ab \notin E$, $t_ac \notin E$, and $t_ca \notin E$, $t_cb \notin E$. Since $b$ is white, $a$ and $c$ are black, and there are black $M$-mates $a' \in N_4$ of $a$ and $c' \in N_4$ of $c$ (recall that by (\ref{noMedgesN3N4}), there is no $M$-edge between $N_3$ and $N_4$). By Claim~\ref{componinN4}, $a'b \notin E$ and $c'b \notin E$.

Now by Corollary \ref{P4inN4C5} $(i)$, $t_ac' \in E$ and $t_ca' \in E$. Clearly, $t_a$ and $t_c$ are white, and thus, $t_at_c \notin E$. Since $(t_a,a,b,c,t_c)$ do not induce a $P_5$ with $t_c \notin T_1$ (else it leads to a $P_9$ in $G$), we have $t_c \in T_1$.

Suppose that $t_b \notin T_1$. Then, since $(t_b,b,c,c',t_a)$ do not induce a $P_5$ (else there is a $P_9$ in $G$ with midpoint $t_a$), we have $t_bt_a \in E$, and analogously, since $(t_b,b,a,a',t_c)$ do not induce a $P_5$, we have $t_bt_c \in E$ but now, $t_b \notin T_1$ contacts two vertices in $T_1$, which is a contradiction (recall Lemma \ref{lemm:structure2} $(v)$). Thus, $t_b \in T_1$, and Claim \ref{N4P3abcwhiteb} is shown.
\qed

\begin{corollary}\label{N4componP3abcwhiteb}
For a component $D$ in $G[N_4]$ with $P_3$ $(a,b,c)$ such that $b$ is white, there are three $N_3$-neighbors of $D$ in the same $T_i$ such that every vertex of $D$ contacts one of them.
\end{corollary}

\noindent
{\bf Proof.}
If $D$ is a $P_5$ $(a',a,b,c,c')$ as in the proof of Claim \ref{N4P3abcwhiteb} then clearly, there are three $N_3$-neighbors of $D$ in the same $T_i$
such that every vertex $a',a,b,c,c'$ contacts one of them.

Clearly, $D$ is $P_9$-free. Now assume that there is a neighbor $d \in N_4$ of $c'$ (recall that every vertex in $D$ has degree at most 2). Clearly, $d$ is white, $dc \notin E$ and $db \notin E$.
Let $t_b \in N_3$ be a neighbor of vertex $b$.
Since $(t_b,b,c,c',d)$ do not induce a $P_5$, by the discussion at the beginning of the section,
we have $t_bd \in E$. Accordingly, if $e \in N_4$ is a neighbor of $d$ and $t_c \in N_3$ is a neighbor of vertex $c$ then, since $(t_c,c,c',d,e)$ do not induce a $P_5$, we have $t_ce \in E$ etc.
Thus, Corollary \ref{N4componP3abcwhiteb} is shown.
\qed

\begin{clai}\label{D1D22P2awhitecontactcolor}
Let $D_1,D_2$ be two components in $G[N_4]$ and let $a,b \in V(D_1)$ with white vertex $a$ and $ab \in E$ as well as $c,d \in V(D_2)$ with $cd \in E$.
Then $c$ and $d$ are colored black by the white vertex $a$.
\end{clai}

\noindent
{\bf Proof.}
Let $t_a \in T_1$ be an $N_3$-neighbor of $a$. Then $t_a$ is black, and all other vertices in $T_1$ are white. Let $t_c \in N_3$ be a neighbor of $c$.
Clearly, $t_ab \notin E$ and $t_cd \notin E$, and $ab,cd$ induce a $2P_2$ in $G[N_4]$.
If $t_c \notin T_1$, say $t_c \in T_2$, and $t_at_c \notin E$ then $(b,a,t_a,u_1)$, $(d,c,t_c,u_2)$, and the shortest path in $N_1 \cup \{x,y\}$ between $u_1$ and $u_2$ lead to a $P_9$, which is a contradiction. Thus, either $t_c \in T_1$ or $t_ct_a \in E$ which implies that $t_c$ is white, and thus, $c$ is black.
Analogously, $d$ is colored black by the white vertex $a$, and Claim \ref{D1D22P2awhitecontactcolor} is shown.
\qed

\begin{corollary}\label{N4onlyonecomponD1}
There is only one component in $G[N_4]$.
\end{corollary}

\noindent
{\bf Proof.}
Suppose to the contrary that there are two such components $D_1,D_2$ in $G[N_4]$. Clearly, by Claim \ref{N4isolatededgeblack} and the Edge Reduction,
$D_1$ contains a white vertex $a$; let $ab \in E$ for $a,b \in V(D_1)$.
As in the proof of Claim~\ref{D1D22P2awhitecontactcolor}, let $t_a \in T_1$ be an $N_3$-neighbor of $a$ which is black, and all other vertices in $T_1$ are white, and $c$ and $d$ are black for an edge $cd \in E$, $c,d \in V(D_2)$. Clearly, $D_2$ has at least three vertices; let $e$ be a neighbor of $c$ or $d$, say $de \in E$. Then $e$ is white, and thus, an $N_3$-neighbor $t_e$ of $e$ is black, and thus, $t_e \notin T_1$ and $t_at_e \notin E$; let $t_e \in T_2$. But now, $(b,a,t_a,u_1)$, $(d,e,t_e,u_2)$, and the shortest path in $N_1 \cup \{x,y\}$ between $u_1$ and $u_2$ lead to a $P_9$, which is a contradiction.
Thus, Corollary \ref{N4onlyonecomponD1} is shown.
\qed

\medskip

Let $K$ be a nontrivial component of $G[S_2 \cup N_3 \cup N_4]$, and let $Q_1,\ldots,Q_{\ell}$ be the components of $K$ in $G[S_2 \cup N_3]$ and let $D$ be the
component of $K$ in $G[N_4]$. For each of the (at most three) feasible colorings of $D$, it leads to a partial coloring in every $Q_i$ since there is no contact between $Q_i$ and $Q_j$, $i \neq j$, and thus, there are contacts between $Q_i$ and $D$. Then, as in Section \ref{N4empty},
 for every $Q_i$, it can be independently checked in polynomial time whether $Q_i$ has a feasible coloring or a contradiction.

\medskip

This finally shows:

\begin{theorem}\label{DIMP9frpol}
DIM is solvable in polynomial time for $P_9$-free graphs.
\end{theorem}

\section{Conclusion}

In \cite{CarKorLoz2011}, it is shown that for every graph class of bounded clique-width, the DIM problem can be solved in polynomial time.
However, there are many examples where the clique-width is unbounded but DIM is solvable in polynomial time; for example, the clique-width of $P_9$-free graphs is unbounded. The complexity of DIM is still an open problem for many examples.

\medskip

\noindent
{\bf Acknowledgment.}
We are grateful to the anonymous referees for their helpful comments.
The second author would like to witness that he just tries to pray a lot and is not able to do anything without that - ad laudem Domini.

\begin{footnotesize}

\end{footnotesize}

\end{document}